\documentclass[12pt]{article}
\usepackage{xcolor}
\usepackage{cite}
  \usepackage{graphicx}
  \usepackage{epsfig}
  \usepackage{amssymb}
  \usepackage{subfigure}
  \usepackage{physics}
  \usepackage{amsmath} 

\evensidemargin - .5truecm
\allowdisplaybreaks[1]

\begin{document}
\newcommand*\xbar[1]{\hbox{\vbox{%
      \hrule height 0.5pt 
      \kern0.3ex
      \hbox{%
        \kern-0.2em
        \ensuremath{#1}%
        \kern -0.1em
      }}}} 
\newcommand{\lsim}{\lesssim}
\newcommand{\gsim}{\gtrsim}

\newcommand{\CC}{{\mathbb C}}
\newcommand{\RR}{{\mathbb R}}
\newcommand{\ZZ}{{\mathbb Z}}
\newcommand{\QQ}{{\mathbb Q}}
\newcommand{\NN}{{\mathbb N}}
\newcommand{\beq}{\begin{equation}}
\newcommand{\eeq}{\end{equation}}
\newcommand{\beal}{\begin{align}}
\newcommand{\eeal}{\end{align}}
\newcommand{\nn}{\nonumber}
\newcommand{\bea}{\begin{eqnarray}}
\newcommand{\eea}{\end{eqnarray}}
\newcommand{\ba}{\begin{array}}
\newcommand{\ea}{\end{array}}
\newcommand{\bfig}{\begin{figure}}
\newcommand{\efig}{\end{figure}}
\newcommand{\bc}{\begin{center}}
\newcommand{\ec}{\end{center}}

\newenvironment{appendletterA}
{
  \typeout{ Starting Appendix \thesection }
  \setcounter{section}{0}
  \setcounter{equation}{0}
  \renewcommand{\theequation}{A\arabic{equation}}
 }{
  \typeout{Appendix done}
 }
\newenvironment{appendletterB}
 {
  \typeout{ Starting Appendix \thesection }
  \setcounter{equation}{0}
  \renewcommand{\theequation}{B\arabic{equation}}
 }{
  \typeout{Appendix done}
 }

\begin{titlepage}
\nopagebreak

\renewcommand{\thefootnote}{\fnsymbol{footnote}}
\vskip 2cm

\vspace*{1cm}
\begin{center}
{\Large \bf 
  Saddle-point method for
  \\[0.1cm]
  resummed form factors in QCD
}
\end{center}


\par \vspace{1.5mm}
\begin{center}
    {\bf Ugo Giuseppe Aglietti${}^{(a)}$},  
    {\bf Giancarlo Ferrera${}^{(b)}$}
and
    {\bf  Wan-Li Ju${}^{(c)}$} 
\vspace{5mm}

${}^{(a)}$
Dipartimento di Fisica, Universit\`a di Roma 'La Sapienza' \\ and INFN, Sezione di Roma
I-00185 Rome, Italy\\\vspace{1mm}

${}^{(b)}$ 
Dipartimento di Fisica, Universit\`a di Milano and INFN, \\  Sezione di Milano,
I-20133 Milan, Italy\\\vspace{1mm}

${}^{(c)}$ 
Department of Physics, University of Alberta, Edmonton \\ AB T6G 2J1, Canada

\end{center}

\vspace{.5cm}


\par 
\begin{center} {\large \bf Abstract} \end{center}
\begin{quote}
\pretolerance 10000

We consider the form factor appearing in QCD resummation formalism for event shape
distributions in the two-jet (or Sudakov) region.
We present an analytic formula for the 
inverse transform of the form factor,
namely from the conjugate moment space 
to the (physical) momentum space, 
based on the saddle-point method.
The saddle-point itself is determined
by means of an analytic recursion method
as well as by standard numerical methods.
The results we have found are in very good agreement
with the exact (numerical) evaluation
of the inverse transform, 
while they significantly differ from classical analytical formulations
of resummation in momentum space.
The latter are
based on a Taylor expansion of the form factor around the free-theory 
saddle point.

\vskip .4cm

\vfill\end{quote}

\begin{flushleft}
June 2025
\vspace*{-1cm}
\end{flushleft}
\end{titlepage}

\renewcommand{\thefootnote}{\fnsymbol{footnote}}

\newpage


\section{Introduction}

A fully analytic formalism for the resummation
of $e^+e^-$ event shape distributions in the two-jet (or Sudakov) region 
in physical
space is a long-standing problem
in perturbative QCD.
The resummation of infrared logarithms (of soft and collinear origin) 
to all orders in perturbation theory is indeed 
naturally accomplished in conjugate (moment)
spaces.
In the latter, multi-parton kinematical constraints factorize 
single-particle ones in the soft limit, 
unlike what happens in space.
Phase space factorization is essential to {\itshape exponentiate}, and thus to resum to all orders, the large infrared logarithms 
contained in the resummed form factor.
The latter is evaluated in physical space
by means of an inverse transform.
However, the (1-dimensional) integral defining
such transform is too
complicated to allow for an exact analytic evaluation.
Nevertheless, such (improper) integral 
can be easily evaluated numerically basically with arbitrary precision.
In any case, obtaining
an accurate analytic approximation of the
resummed form factor in physical space is still
desirable.

A classic solution to this problem
was provided long time ago by the seminal work of Catani, Trentadue, Turnock and Webber
(CTTW)\,\cite{Catani:1992ua}, 
which presented
an approximate analytic formula,
relatively simple, for the
QCD resummed form factor in (physical) momentum space.
The approximate inversion provided by CTTW was later reformulated in the framework of the Soft Collinear Effective Theory (SCET)\,\cite{Bauer:2008dt,Almeida:2014uva}.
This solution has been the standard method for the subsequent implementations of Sudakov resummation in shape variables
both in QCD and in SCET\,\cite{Gardi:2001ny,Banfi:2001bz,Schwartz:2007ib,Becher:2008cf,Abbate:2010xh,Monni:2011gb}. 

Recently, in Ref.\,\cite{Aglietti:2025jdj}, the resummation of thrust  distribution\,\cite{Farhi:1977sg} has been investigated.
It was found that the spectra obtained by using the approximated analytic formula in momentum space
substantially differ from the spectra
obtained by means of an exact (numerical) inversion within the so-called Minimal Prescription
of Ref.\,\cite{Catani:1996yz}.
In particular, the differences 
in the calculated spectra
have been found to be much larger than the related perturbative uncertainties.
It has been pointed out that such differences originate from approximations of the analytic inversion formula, whose impact is usually neglected in the literature.
An important phenomenological consequence 
of this fact is that, in the case of the thrust, the commonly used {\itshape momentum}-space formulation
produces a determination of 
$\alpha_S(m_Z^2)$ not consistent with the current world average, in contrast with the case of exact numerical inversion which is
fully consistent with the latter \,\cite{Aglietti:2025jdj}.

In this paper, we analyze in detail the problem of the analytic inversion of the resummed form factors for shape variables and propose a novel analytic solution based again on the
saddle-point method. 
The main feature of our approach is 
that of Taylor expanding the exponent of the integrand around its ``true'' saddle-point, that is, the saddle point of 
the full (interacting) theory.
On the contrary, the standard method is based on an expansion around the free-theory saddle point.
By comparing our approximate analytic inversion with the exact numerical one, we will show that the our formalism provides more accurate results than the standard method.
Moreover, we are able to confirm the numerical results of Ref.\,\cite{Aglietti:2025jdj}.

As far as we know, "true" saddle-point methods have never been  applied for the analytic inversion of the resummed form factor in the case of shape variables.
Saddle-point expansion in the related cases of threshold and transverse-momentum resummations have been treated in Refs.\,\cite{Bonvini:2012an,Bonvini:2014qga,Kang:2017cjk,Grewal:2020hoc}.

The paper is organized as follows.
In Sec.\,\ref{sec_formalism}, we briefly
review the classical CTTW formalism for the resummation of event shape variables.
In Sec.\,\ref{sec_saddle}, we discuss 
the standard (mathematical) saddle-point method
and we apply it to the case 
of the inverse transform
of the form factor occurring in event shape resummation.
The "true" saddle point
is the solution of a (very complicated) 
transcendental functional equation.
We determine it both  numerically (basically with arbitrary precision), 
by standard methods, and
analytically, by means of 
a recursive method.
The application 
of the saddle-point method to resummed form factors is made
complicated by
the occurrence of the well-known
(unphysical) Landau singularity of the QCD coupling in the infrared region.
We regularize the resummed form factor 
by replacing the standard (infrared divergent) QCD coupling
with an analytic effective coupling free from the Landau singularity\,\cite{Aglietti:2004fz}.
In Sec.\,\ref{sec_pheno}, we present our numerical results for the
regularized resummed form factor 
for the case of the thrust.
We compare our results with 
the exact numerical ones
as well as with the classical ones.
Finally, in App.\,\ref{sec_frozen_coupling}, we consider the frozen coupling limit for the resummed form factor.
In this limit,
the form factor simplifies 
to such an extent that we are able to
identify the (large) expansion
parameter formally controlling 
the convergence of the saddle-point expansion.


\section{Event-shape resummation}
\label{sec_formalism}

The resummation of $e^+e^-$ event-shape distributions in the two-jet (or back-to-back) region has been formalized in a seminal paper by CTTW in\,\cite{Catani:1992ua}, which we
briefly review. A generic infrared 
(i.e. soft and collinear) safe distribution, depending on the dimensionless event-shape variable $y$ vanishing in the two-jet limit (or Sudakov
region), has the following perturbative expansion in $\alpha_S=\alpha_S(Q^2)$%
\,\footnote{
$Q$ is the hard scale, i.e. the reference energy scale of the hard process.
This scale has to be much larger than the QCD scale $\Lambda_{QCD}$ ($Q\gg \Lambda_{QCD}$)
for the application of perturbative QCD.
In $e^+e^-$ annihilation to hadrons,
$Q$ is typically the center-of-mass
energy of the $e^+e^-$ pair. 
}
at small (but not vanishing) $y$:
\beq
\label{diff}
\frac1{\sigma}\frac{d\sigma}{dy}= 
\sum_{n=1}^\infty \sum_{m=0}^{2n-1} \, c_{n,m} \, \alpha_S^n\,
\frac{\ln^{m}y}{y} \, + \, \cdots \,,
\eeq
where $\sigma$ is the total cross section of $e^+e^-$ annihilation into hadrons, $c_{n,m}$ are $y$-independent numerical coefficients, 
$\ln^{m}y/y$ are the so-called Sudakov 
logarithms of soft and collinear origin
enhanced at small $y$.
The neglected terms are remainder contributions less singular for $y\to 0$. In the small-$y$ region ($y\ll 1$), when $\alpha_S \ln(y)/y \sim 1$, the fixed-order
perturbative expansion is not reliable. In this region, the resummation to all orders of the enhanced logarithmic terms is necessary in order to obtain accurate perturbative predictions.

The resummation of the Sudakov logarithms, with the inclusion of the virtual contributions at the $y=0$ endpoint, is conveniently performed by considering the {\itshape cumulative} distribution
\beq
\label{cumu}
R(y) \, \equiv \, \frac1{\sigma} \int_0^y dy' \frac{d\sigma}{dy'}\,. 
\eeq
In this paper we deal with event-shape variables for which, up to some level of accuracy, the large logarithmic contributions {\itshape exponentiate}. That means that
the factorization properties of QCD matrix elements {\itshape and} of the phase space 
constraints
defining $y$  in the two-jet region  allow the following
factorized expression for the cumulant cross section $R(y)$\,\cite{Catani:1992ua}:
\begin{equation}
\label{eq:cumulant_expansion}
    R(y) =  C\left(\alpha_S\right)\Sigma\left(y,\alpha_S\right) + D\left(y,\alpha_S\right).
\end{equation}

$C(\alpha_S)$ is a hard-virtual factor with a standard (fixed-order) perturbative expansion
\begin{equation}
\label{Chard}
   C(\alpha_S)= 1+ \sum_{n=1}^{\infty} \left(\frac{\alpha_S}{\pi}\right)^n C_n\,;
\end{equation}
where the $C_n$'s are numerical coefficients.

$D(\alpha_S)$ is a short-distance {\itshape remainder} function, vanishing at small $y$,
\begin{equation}
\label{Drem}
   D(y,\alpha_S)= \sum_{n=1}^{\infty} \left(\frac{\alpha_S}{\pi}\right)^n D_n(y)\,.
\end{equation}

$\Sigma\left(y,\alpha_S\right)$ is a long-distance dominated form factor, that contains (and resums) all the logarithms  enhanced at small $y$.
As already stated, the resummation of Sudakov logarithms is feasible if the form factor possesses factorization properties and can be recast in an {\itshape exponential} form, so that
multiple soft-gluon corrections are generated by iterations
of single soft-gluon emission.

Typically, phase-space factorization, and thus exponentiation, does not hold in the physical, momentum space in which the variable $y$ is defined, but in conjugate-moment ($N$) space. In such
space, the $N$-moment of the factorized form factor $\widetilde\Sigma_N(\alpha_S)$ can be written as:
  \begin{equation}\label{ffactor}
    \widetilde\Sigma_N(\alpha_S) =\exp[ {\mathcal F}(\alpha_S,L)]
    =\exp\left[ L\,f_1(\lambda)   + \sum_{n=2}^{\infty} \left(\frac{\alpha_S}{\pi} \right)^{n-2} f_n(\lambda)\right]\,,
  \end{equation}
    where $L\equiv\ln N$, $\lambda\equiv\beta_0\alpha_S L/\pi$ and $\beta_0$ is the first-order coefficient of the QCD $\beta$ function. 
  The exponent ${\mathcal F}(\alpha_S, L)$  in Eq.\,(\ref{ffactor})
  resums to all orders  in $\alpha_S$ series of logarithms of $\ln N$, which are large for $N\to \infty$ (and approximately correspond, in physical space, to
  $\ln y$ terms enhanced in the two-jet region $y \to 0$).
  At large $N$, $L \gg 1$ and $\alpha_S L$ is assumed to be $\mathcal{O}(1)$,
  implying that the exponent in the r.h.s.\ of Eq.\,(\ref{ffactor}) has a customary perturbative expansion in powers of $\alpha_S$.
The truncation of such a function series at a given order yields the resummation of towers of 
logarithmic corrections.  
The inclusion of the functions $f_{n}(\lambda)$ up to $n=k+1$ corresponds to the $k$-order logarithmic  (N$^{k}$LL) accuracy.

The form factor in physical space, $\Sigma\left(y,\alpha_S\right)$, can then be obtained by the inverse transform of $\widetilde\Sigma_N(\alpha_S)$ from moment space $N$ to momentum space $y$.
However, such inverse transform is too complicated to be performed {\itshape exactly} in closed analytic form even at lowest-order or leading-logarithmic (LL) accuracy.
Moreover, the expression of the form factor $\widetilde\Sigma_N(\alpha_S)$ given in Eq.\,(\ref{ffactor}) only has a formal meaning, as it
suffers from the Landau singularity of the QCD running coupling,
that manifests itself in
singularities of the $f_n(\lambda)$ functions at very large values of $N$ (i.e.\ in momentum space, in the region of very small $y$). 
These singularities, which signal the onset of non-perturbative (NP)
phenomena at very small momenta, 
have to be properly regularized through prescriptions or non-perturbative models
to render the inverse transform of  $\widetilde\Sigma_N(\alpha_S)$ feasible.

Two main alternatives for the computation of $\Sigma(y,\alpha_S)$ are typically followed in the literature. 
The first alternative, originally developed in\,\cite{Catani:1992ua}, 
is an analytic approximate calculation of the form factor
based on a Taylor expansion around the point $N=1/y$ where 
\begin{equation}
\label{Nwrong}  
\ln N=  \ln(1/y)\equiv \ell \,,
\end{equation}
which gives
\begin{equation}\label{ffactor_taylor}
  \widetilde\Sigma_N(\alpha_S)=  
 \exp\!\left[\sum_{k=0}^\infty\frac{{\mathcal F}^{(k)}(\alpha_S,\ell)}{k!}\ln^k(y N)\right]\,, 
\end{equation}
where 
${\mathcal F}^{(k)}(\alpha_S,\ell)\equiv{\partial^{k}}{\mathcal F}(\alpha_S,\ell)/{\partial\ell^{k}}\,.$
Since it is not possible to evaluate  the inversion transform of the series on the r.h.s.\ of  Eq.\,(\ref{ffactor_taylor})
exactly, a hierarchy is defined in $y$ space. 
The N$^n$LL accuracy in $y$ space is defined by
keeping in Eq.\,(\ref{ffactor_taylor}) only the dominant logarithmic terms $\alpha_S^{n-1}(\alpha_S\ell)^k$,
up to a given $n$ and for all $k$, and performing the inverse transform analytically.
However, it is important to stress that the resulting resummation formula in $y$ space is 
only an approximation of the resummation formula in $N$ space. 
In particular, the resummation in $y$ space at a given logarithmic accuracy
{\itshape does not} resum all the $\ln N$ terms
of the corresponding logarithmic accuracy in $N$ space.
The crucial point is that there is no exact correspondence
between $\ln N$ and $\ln(1/y)$ terms. 
This is because the kinematical constraints of momentum conservation  factorize in $N$ space, not in $y$ space, and thus exponentiation
is strictly valid only in $N$ space.
We also point out that within this approach an {\itshape implicit} regularization prescription for the Landau singularity is assumed, otherwise
the analytic inversion would not be feasible. Since the analytic inversion from $N$ to $y$ space is typically performed by means of the residue theorem,
the prescription consists of neglecting the residue due to the Landau singularity. Nevertheless, this procedure is fully justified from the point of view of perturbation theory since
the ambiguities due to the Landau singularity  give rise to power-suppressed contribution that are always missed
in a perturbative expansion\,\cite{Catani:1996yz}.

The second alternative is to perform the inverse transform {\itshape exactly} by numerical integration. Of course, also in this case, explicit prescriptions to
avoid or regulate the Landau singularity are necessary\,\cite{Catani:1996yz,Forte:2006mi,Aglietti:2006yb,Aglietti:2006yf}.
In Ref.\,\cite{Aglietti:2025jdj}, we have recently shown, in the case of thrust resummation, that
the spectra obtained with the approximated analytic expansion and the exact numerical inversion using the so-called Minimal Prescription\,\cite{Catani:1996yz}
to regulate the Landau singularity substantially differ.
In particular the differences have been found to be larger than  the perturbative uncertainties
estimated both through perturbative scale variation and through the difference between two consecutive perturbative orders.
We pointed out that these differences originate from the approximations of the analytic inversion formula whose impact has been usually neglected in the literature.
An important consequence is that resummation formalism based on the commonly used approximated analytic  
inversion gives, contrary to the case of the exact numerical inversion, a corresponding lower determination of $\alpha_S(m_Z^2)$ from the thrust distribution
that is not consistent (within uncertainty) with the world average.

The form factor $\widetilde\Sigma_N(\alpha_S)$ and its inverse transform $\Sigma(y,\alpha_S)$ are the main objects which we analyze in the present paper.
In particular, using standard saddle-point methods, which date back to Laplace\,\cite{Laplace1814}, we show that it is possible to obtain an accurate analytic approximation of the form
factor $\Sigma\left(y,\alpha_S\right)$
that fully agrees, within uncertainties, with the result obtained
by exact numerical inverse transform.


\section{Saddle-point method for resummed form factors}
\label{sec_saddle}

We consider the form factor of the cumulative 
event-shape distribution $y$  in Eq.\,(\ref{ffactor}).
In particular, we analyze the explicit case of the thrust distribution, $y=\tau$,
but the results obtained are generic for the resummation of other shape-variables (such as heavy-jet mass or $C$-parameter)
and also for the case of threshold resummation. This is because the structure of the form factor in Eq.\,(\ref{ffactor}) is the same.

The form factor $\Sigma(y,\alpha_S)$ is obtained via the
inverse Laplace transform
of the distribution in $N$-space:
\beq
\label{eq_Sudakov_cumulative}
\Sigma(y,\alpha_S) =  \frac1{2 \pi i} \int_{\mathcal C} 
\frac{dN}{N}
e^{y N}\, \widetilde\Sigma_N(\alpha_S)  
= \frac1{2 \pi i} \int\limits_{c-i\infty}^{c+i\infty}\,dN
\exp[ g_y(N) ]\,
\eeq
where the function $g_y(N)$ at the exponent is given by (see Eq.\,(\ref{ffactor})):
\beq
\label{gy}
g_y(N) \, \equiv \, y \, N \, - \, \ln N +
\ln(N) \, f_1(\lambda) \, + 
\sum_{n=0}^\infty \left(\frac{\alpha_S}\pi\right)^n \, f_{n+2}(\lambda).
\eeq
The contour $\mathcal C$ in the complex $N$ plane must be chosen such that all singularities of the integrand lie to its left and the contour extends to infinity. For instance
$\mathcal C$ can be a line parallel to the imaginary axes and passing by the real constant $c$ 
(the actual value of $c$ is irrelevant provided the condition above is fulfilled).

The type of integral in  Eq.\,(\ref{eq_Sudakov_cumulative}) can be performed by applying standard saddle-point methods as follows.
Given an analytic function $h(\nu)$ and a large parameter $s\gg 1$ we have:
\bea
\label{eq_saddle_point}
\int_{-\infty}^{+\infty} \frac{d\nu}{2\pi} \varphi(\nu) 
\exp\left[ -s \, h(\nu) \right]
&=&
\int_{-\infty}^{+\infty} \frac{d\nu}{2\pi} \varphi(\nu) 
\exp\left[ -s \, h(\bar\nu) - s\, h''(\bar{\nu}) \frac{\nu^2}2-s\,\sum_{k=3}^{\infty}  h^{(k)}(\bar{\nu}) \frac{\nu^k}{k!} \right]\nn\\
&=&  \varphi(\bar{\nu})
\frac{ \exp\big[ -s \, h(\bar{\nu}) \big] }{ 
  \sqrt{2 \pi s  h''(\bar{\nu})} }
\bigg[1  +  \mathcal{O}\bigg( \frac{1}{s} \bigg)
\bigg]\,,
\eea
where $\bar{\nu}$ is the (unique) minimum of the function $h(\nu)$, so that  
\beq
h'(\bar{\nu}) \, = \, 0\,,
\eeq
with (generic case) 
\beq
h''(\bar{\nu}) \, > \, 0.
\eeq
In the last equality of Eq.\,(\ref{eq_saddle_point}), we have analytically performed the quadratic (harmonic) Gaussian integral and neglected the (anharmonic) higher-order 
corrections, that are suppressed by inverse powers of the large parameter $s$.

The saddle-point approximation in Eq.\,(\ref{eq_saddle_point}) can be applied
to the form factor in Eq.\,(\ref{eq_Sudakov_cumulative}) by taking $c=\xbar{N}>0$
and using the following parametrization
of the integration variable: 
\beq
N \, = \, \xbar{N} \, + \, i \, \nu \qquad (\nu \in \RR)\,,
\eeq
where $\xbar{N}$ is the saddle point for  $g_y(N)$.
Eq.\,(\ref{eq_Sudakov_cumulative}) is then explicitly rewritten:
\beq
\label{eq_Sudakov_cumulative_2}
\Sigma(y,\alpha_S) \, = \, \int\limits_{-\infty}^{+\infty}\frac{d\nu}{2\pi}
\exp\big[ h_y(\nu) \big]\,,
\eeq
where we have defined the function of the real variable $\nu$:
\beq
h_y(\nu) \, \equiv \, g_y(\,\xbar{N} + i \nu)\,.
\eeq

The saddle-point equation for $\xbar{N}=\xbar{N}(y)$,
\beq
g_y'(\,\xbar{N}\,)  =  0\,,
\eeq
explicitly reads
\beq
\label{eq_find_saddle_point_expl}
y \, \xbar{N}  =  1 
 -  \frac{d}{d\lambda} \left[  
\lambda \, f_1(\lambda) 
 +  \beta_0 \sum_{n=1}^{\infty}
\left(\frac{\alpha_S}\pi\right)^n
f_{n+1}(\lambda)
\right]_{\lambda \, \mapsto \, \bar \lambda }\,,
\eeq
where $\bar{\lambda} \equiv \beta_0\alpha_S \ln(\,\xbar{N}\,)/\pi$.
Applying the saddle-point formula, we then obtain the following analytic expression for the form factor in $y$ space:
\beq
\label{harmonic}
\Sigma(y,\alpha_S)  \simeq  
\frac{ \exp\big[ h_y(\nu=0) \big] }{ 
\sqrt{2 \pi \, h_y''(\nu=0)}}\,,
\eeq
where the function $h_y(\nu)$ has its minimum at $\nu=0$ (i.e.\ at $N=\xbar{N}$) and
\beq
h_y''(\nu=0)=-g_y''(\,\xbar{N}\,)>0\,.
\eeq

By comparing the r.h.s.\ of Eq.\,(\ref{eq_Sudakov_cumulative_2}),
with the integral involved in the
saddle-point approximation, on the l.h.s.\ of
Eq.\,(\ref{eq_saddle_point}), 
we note in the former the absence of the control parameter $s$.
We also observe that this expansion parameter is necessary to formally control the $\mathcal{O}(1/s)$ (anharmonic) higher-order corrections
to the Gaussian approximation, which are small 
only for $s \gg 1$.
In our case, the parameter
$s$ is embodied in the function $h_y(\nu)$, the exponent of the integrand function, 
and cannot be extracted out of the latter.
That implies that there is no proof that 
our (saddle-point inspired) approximation of the integral in Eq.\,(\ref{eq_Sudakov_cumulative_2}) is valid for any value of $y$.
In practice, the accuracy of the saddle-point approximation depends on the fast decay of  the function $h_y(\nu)$ around its maximum. 
That implies an exponential suppression of 
the integrand $\exp{h_y(\nu)}$ away from its 
maximum at $\nu=0$
(a saddle point upon analytic continuation,
hence the generic name saddle-point method).

However, in order to check the accuracy of the saddle-point expansion, it is possible to calculate the  {\itshape anharmonic} corrections
to the Gaussian approximation, which read
\beq
\label{anharm}
\Sigma(y,\alpha_S) = 
\frac{ \exp\big[ h_y(\nu=0) \big] }{ 
\sqrt{2 \pi \, h_y''(\nu=0)} }\int_{-\infty}^{+\infty}\frac{dx}{\sqrt\pi}\,e^{-x^2}\,K(x)\,,
\eeq
with:
\beq
\label{kappa}
K(x) \, = \, \exp[\sum_{k=3}^{\infty}\,c_k\,(ix)^k]\,,
\eeq
where we have defined the coefficients
\beq
c_k \, = \, \frac{h_y^{(k)}(\nu=0)}{k!}\left[\frac{2}{h_y^{''}(\nu=0)}\right]^{k/2}\,.
\eeq
In Eq.\,(\ref{anharm}) we have made the change of variable $x=\nu\,\sqrt{h_y''(\nu=0)/2}$.
Let us also note that, if Eq.\,(\ref{kappa}) is expanded in power series
the odd terms in $x$ vanish, upon integration over $x$, by parity
(without such expansion, their contributions are suppressed
but non-vanishing).
Moreover, let us observe that the convergence of the integral in the r.h.s.\ of Eq.\,(\ref{anharm})
depends on the signs of the coefficients $c_k$ and therefore, the anharmonic
terms can be included only if $c_k i^k<0$.
Finally, we stress that, in general, anharmonic corrections in the r.h.s.\ of Eq.\,(\ref{anharm}) cannot be calculated exactly in closed analytic form as the dominant Gaussian
term in Eq.\,(\ref{harmonic}) is.

The application of the saddle-point method to the form factor inversion is thus simply reduced to solving the saddle-point Eq.\,(\ref{eq_find_saddle_point_expl})
and then  applying the formula in Eq.\,(\ref{harmonic}).
The saddle-point Eq.\,(\ref{eq_find_saddle_point_expl}) can be solved either by means of standard numerical methods or analytically by means of a recursive solution as explained below.


\subsection{Recursive solution of saddle-point equation}

In order to explicitly evaluate 
the saddle-point 
$\xbar{N}$ by solving 
Eq.\,(\ref{eq_find_saddle_point_expl}),
it is convenient to introduce, 
in place of $\xbar{N}=\xbar{N}(y)$, the variable $u=u(y)$,
defined as:
\beq
u \, \equiv \, y \, \xbar{N} \, > \, 0\,.
\eeq
In the new variable, the saddle-point Eq.\,(\ref{eq_find_saddle_point_expl}) is rewritten:
\beq
\label{eq_saddle_point_in_y}
u  =  \Phi(u)\,,
\eeq
where we have introduced the function:
\beq
\Phi(u)  \equiv 
1 
 - \frac{d}{d\lambda} \left[  
\lambda \, f_1(\lambda) 
 +  \beta_0 \sum_{n=1}^{\infty}
\left(\frac{\alpha_S}{\pi}\right)^n
f_{n+1}(\lambda)
\right]_{\lambda \, \mapsto \, \bar\lambda },
\eeq
where $\bar\lambda = \alpha_S\beta_0\ln(\xbar{N})/\pi=\alpha_S\beta_0\ell/\pi+\alpha_S\beta_0\ln(u)/\pi$\,, (we recall, see Eq.\,(\ref{Nwrong}), that $\ell=-\ln y)$.
The above equation, involving nested logarithms,
cannot be solved in closed
analytic form in terms of (known) transcendental functions.
Indeed, already at the LL accuracy
Eq.\,(\ref{eq_saddle_point_in_y}) explicitly reads\,\cite{Catani:1991kz}: 
\beq
\label{spLL}
u = 
1  +
\frac{2 A_1}{\beta_0} 
\ln\left[
\frac{1 -  \alpha_S\beta_0 \, (\ell+\ln u)/\pi}{ 
1 - 2 \alpha_S\beta_0 \, (\ell+\ln u)/\pi} 
\right].
\eeq
We thus solve analytically 
Eq.\,(\ref{eq_saddle_point_in_y}), by recursion, as follows:
\bea
u_{n+1} \, = \, \Phi\big( u_n \big)\,,
\qquad n \, = \, 0, 1, 2, \cdots\,,
\eea
with
\beq
u \, = \, \lim_{n\to \infty} u_n\,.
\eeq

The initial condition
(zero-order term) is given by:
\beq
u_0 \, = \, 1.
\eeq
It is possible to solve explicitly
for the $n$-th recursion, that 
is simply written as 
\beq
u_n \, = \, \Phi^{[n]}(u_0=1)\,,
\eeq
where $\Phi^{[n]}(u)$ is the function
$\Phi(u)$ composed with itself 
$n$ times:
\beq
\Phi^{[n]}(u) \, \equiv \, \Phi\Big\{ \Phi\big[\cdots \Phi(u)\big] \Big\}
\qquad ( n \,\, \mathrm{times} ).
\eeq
For example:
\beq
\Phi^{[2]}(u) \, \equiv \, 
\Phi\big[\Phi(u)\big].
\eeq
The first recursion $(n=1)$ explicitly reads:
\beq
\label{eq_first_recur_symb}
u_1  =  
\Phi\left( u_0 = 1 \right)
 = 
1   -  \frac{d}{d\lambda} \left[  
\lambda \, f_1(\lambda) 
 +  \beta_0 \sum_{n=1}^{\infty}
\left(\frac{\alpha_S}{\pi}\right)^n
f_{n+1}(\lambda)
\right]_{\lambda \, \mapsto \, \beta_0\alpha_S \ell/\pi }.
\eeq
At LL accuracy, only the function $f_1(\lambda)$ contributes:
\beq
\label{u1LL}
u_1  = 
1  +  \frac{2 A_1}{\beta_0} 
\ln\left(
\frac{1 -  \alpha_S\beta_0 \, \ell/\pi}{ 
1 - 2 \alpha_S\beta_0 \, \ell/\pi} 
\right)\,,
\eeq
where, as usual, we assume
$\ell \, \gg \, 1$ and $\alpha_S \, \ell \, \sim \, 1$,
%
which implies that the first recursion
$u_1$,
given in Eq.\,(\ref{eq_first_recur_symb}),
contains a (large) $\mathcal{O}(1)$
correction compared to the zero-order one
$u_0=1$.
Therefore, in order to obtain a non-trivial
saddle point, at least
one recursion has to be made, even at LL. In fact, $u=1$ (i.e.\ $\xbar{N}=1/y$) is the saddle-point of the ``free'' theory ($\alpha_S=0$) only.

\begin{figure}[ht]
\begin{center}
\includegraphics[width=0.65\textwidth]{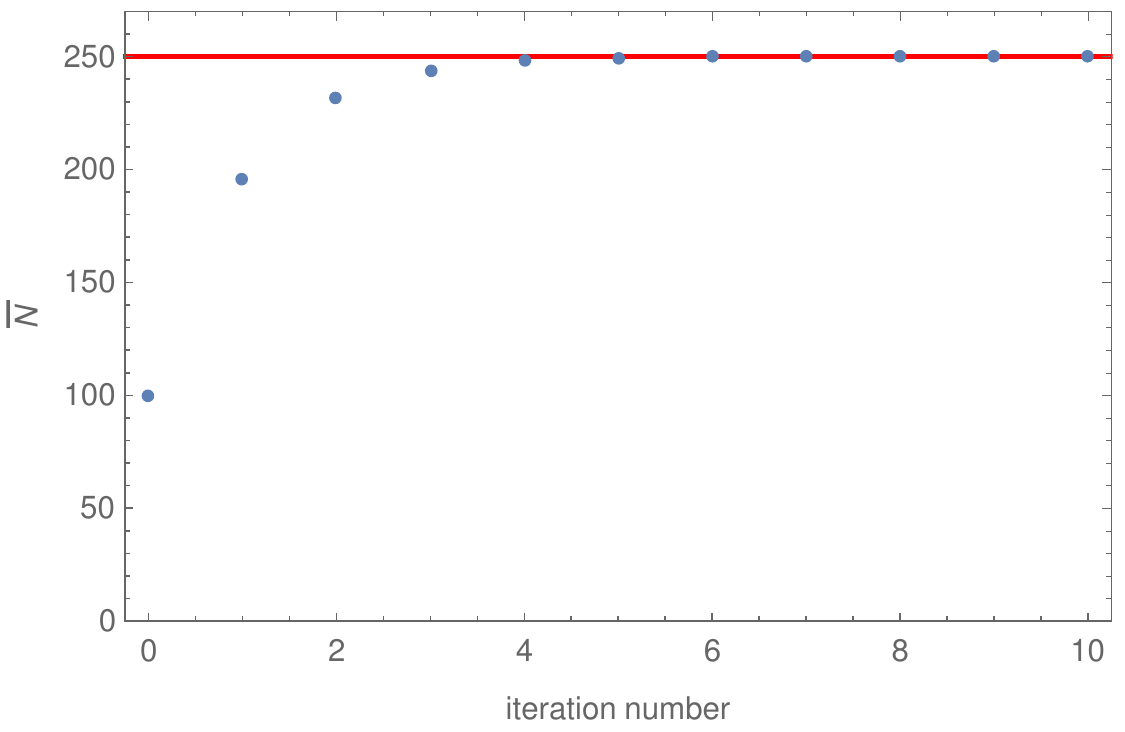}
\end{center}
\vspace*{-.5cm}
\caption{
\label{figSP}
      {\em
Saddle-point solutions at leading logarithmic accuracy (Eq.\,(\ref{spLL})) for the thrust at $y=\tau=0.01$.
   Numerical comparison of the first ten recursions
$\xbar{N}=u_1/y,u_2/y,u_3/y,\cdots$ (blue dots) with the exact numerical solution $\xbar{N}=u/y$ (red line).        
}}
\end{figure}

In order to show the convergence of our recursive sequence $\left\{u_n\right\}$ to the "true"
solution $u$,
in Fig.\,(\ref{figSP}) we show the first ten recursions
$u_1/y,u_2/y,u_3/y,\cdots$ (blue dots),
together
with the exact solution $\xbar{N}=u/y$ (red line), the latter obtained by means of numerical methods, in the case of the thrust form factor at LL for $y=\tau=0.01$. The accuracy of the iterative solution
is better than: $1\%$ for $n=4\,$, $0.1\%$ for $n=6\,$, $0.01\%$ for $n=8$ and $0.0001\%$ for $n=10\,$.


\subsection{Resummed form factor free from Landau singularity}
\label{subsec_reg}

The straightforward application of the saddle-point approximation to the standard form factor in Eq.\,(\ref{ffactor}) is complicated by the presence of the Landau singularity.

Indeed, the LL and NLL functions read: 
\bea
\label{f12}
 f_1(\lambda) & = & \frac{A_1}{\beta_0 \, \lambda}
\Big[
2(1 \, - \, \lambda) \ln(1 \, - \, \lambda)
\, - \, (1 \, - \, 2 \lambda) \ln(1 \, - \, 2 \lambda)
\Big]\,,\\
     f_2(\lambda)& = & \frac{A_2}{\beta_0^2}\Big[\ln(1-2\lambda)-2\ln(1-\lambda)\Big]
  +\frac{2\gamma_E  A_1}{\beta_0} \Big[\ln(1-2\lambda) -\ln (1-\lambda)\Big] 
\nonumber\\  &+&\frac{A_1 \beta_1}{\beta_0^3} \left[\ln ^2(1-\lambda)-\frac12\ln^2(1-2\lambda) -\ln(1-2\lambda)+2\ln(1-\lambda) \right]%
\nonumber\\    &+&\frac{B_1}{\beta_0}\ln (1-\lambda)%
\,,
\eea
where
\begin{align}
      A_1 &=C_F \,,\qquad    A_2 = \frac{C_F}{2}  \left[\left(\frac{67}{18}-\frac{\pi^2}{6}\right) C_A-\frac59 n_f\right]\,,\qquad
      B_1=-\frac{3 C_F}{2}\,,\\
\beta_0 &= \frac{11 C_A \, - \, 2 n_f}{12}\,, 
\qquad\qquad \beta_1 = \frac{1}{24} \left( 17 C_A^2 - 5 C_A n_f - 3 C_F n_f \right)\,, 
\end{align}
with $C_F = 4/3$, $C_A = 3$ and $n_f$ is the number of QCD active
(effective massless) flavors at the hard scale.
The functions  $f_1(\lambda)$ and $f_2(\lambda)$ in Eq.\,(\ref{f12}) are singular at the points
$\lambda = 1/2$ and $\lambda = 1$ (i.e.\ $N\sim N_L= \exp{\pi/[2\beta_0 \alpha_S(Q^2)]}$
and $N\sim N_L' = \exp{\pi/[\beta_0 \alpha_S(Q^2)]}$, respectively). The explicit expressions of the
higher-order logarithmic functions  $f_n(\lambda)$ up to $n=5$ can be found in Ref.\,\cite{Aglietti:2025jdj}.
Similar (actually stronger) singularities appear at higher orders 
and signal the onset of non-perturbative corrections at very large values of $N$ (or, equivalently, in the very small $y$ region).

As already discussed, because of the Landau singularity, the Laplace inversion of the standard resummed form factor, formally does not exist without an explicit or implicit regularization of the latter.
However, it is important to note that the particular method to regularize the infrared region is not relevant to our purpose, since we are
interested in the accuracy of the analytic inversion of the resummed form factor once it is properly regularized.

In this paper, we consider an infrared prescription of the functions $f_i(\lambda)$
where the singularities at $\lambda=1/2$
and $\lambda=1$ have been properly regularized through the procedure proposed in Ref.\,\cite{Aglietti:2004fz} based
on the analytic QCD coupling\,\cite{Shirkov:1996cd,Shirkov:1997wi}.
The analytic coupling at LO reads
\beq
\widetilde\alpha_S(q^2)= \frac1{\beta_0}\left[\frac{\pi}2 - \arctan\frac{\ln(q^2/\Lambda_{QCD}^2)}{\pi}\right]\,. 
\eeq
This coupling is free from the Landau singularity, regular in the infrared region, and reproduces the ultraviolet behavior of the standard one.
Perturbatively, the LO analytic coupling $\widetilde\alpha_S(q^2)$ can be seen as
an all-order expansion of the standard one
\beq
\widetilde\alpha_S(q^2) \, = \, \frac1{\beta_0}\arctan\left({\beta_0 \alpha_S(q^2)}\right)
\, = \, \sum_{n=0}^{\infty}\frac{(-1)^{n}\beta_0^{2n}}{2n+1}\left[\alpha_S(q^2)\right]^{2n+1}\,.
\eeq

The regularized LL and NLL functions obtained by using the analytic time-like
QCD coupling instead of the standard one have been calculated in Ref.\,\cite{Aglietti:2004fz}. 
They explicitly read:
\bea
\label{f1new}
\widetilde{f}_1(\lambda) &=& \frac{A_1}{2\beta_0\lambda}
    \biggl\{ 2(1 - \lambda) \ln[(1 - \lambda)^2+r^2] -( 1 - 2\lambda) \ln[(1-2\lambda)^2+r^2]
     - \ln\left(1+r^2\right)\nn\\
&-& \left. \frac{1}{r}
     \left[
2 \left[( 1 - \lambda)^2-r^2\right] \arctan \left( \frac{1-\lambda}{r} \right)
-\left[( 1 - 2\lambda)^2-r^2 \right] \arctan\left( \frac{1-2\lambda}{r} \right) \right.\right.\nn\\
& -& \left.\left. (1-r^2)\arctan \frac{1}{r} +\pi\lambda^2
\right]
\right\},
\nonumber
\eea
\bea
\label{f2new}
\widetilde{f}_2(\lambda) &=&
\frac{{A_2}} {2 \beta_0^2} \biggl\{ \ln[(1- 2\lambda)^2 + r^2] -2 \ln[(1- \lambda)^2 + r^2] + \ln[1+ r^2]\nonumber \\
&+& \left. \frac{2}{r} \left[- ( 1- 2 \lambda) \arctan\frac{1-2\lambda}{r}+ 2 ( 1-  \lambda) \arctan\frac{1-\lambda}{r} - \arctan\frac{1}{r}\right] \right\} \nonumber \\
&+& \frac{{A_1 \beta_1}}{ 8 \beta_0^3} \biggl\{-\ln^2[(1- 2 \lambda)^2 + r^2] + 2  \ln^2[(1- \lambda)^2 +r^2]-\ln^2[1+ r^2]  \nonumber \\
&+& \left. \frac{2\pi}{r} ( 1-2 \lambda) \left[-1+\frac{2}{\pi}\arctan\frac{1-2\lambda}{r}\right] \ln[(1-2\lambda)^2+r^2]  \nonumber \right. \\
&+& \left. \frac{4\pi}{r} ( 1- \lambda) \left[1-\frac{2}{\pi}\arctan\frac{1-\lambda}{r}\right] \ln[(1-\lambda)^2+r^2]  \nonumber \right. \\
&+& \left. \frac{2\pi}{r}  \left[-1+\frac{2}{\pi} \arctan\frac{1}{r}\right]\ln[1+r^2] +4 \arctan \frac{1-2\lambda}{r}\left[-\pi+\arctan\frac{1-2\lambda}{r}\right]\right. \nonumber \\
&+& \left. 8 \arctan\frac{1-\lambda}{r}\left[\pi- \arctan\frac{1-\lambda}{r}\right] + 4 \arctan \frac{1}{r}\left[-\pi+\arctan\frac{1}{r}\right] \right\}  \nonumber \\
&+& \frac{{A_1 \gamma_E}}{ \beta_0}\biggl\{ \ln[(1- 2 \lambda)^2 + r^2] - \ln[(1-  \lambda)^2 +r^2] \nonumber \\
&+ &  \left. \frac{2}{r} \left[- ( 1- 2\lambda) \arctan\frac{1-2 \lambda}{r}+
  ( 1-  \lambda) \arctan\frac{1-\lambda}{r}-\frac{\pi \lambda}{2}  \right] \right\}\nonumber\\
&+&\frac{{B_1}}{\beta_0}    \bigg\{ \ln [(1 - \lambda)^2+r^2]-\ln [1+r^2] \nonumber\\
&+&  \frac{1}{r} \left[-\pi\lambda+ 2\arctan\frac{1}{r} -2 ( 1 -\lambda )  \arctan \frac{1-\lambda}{r}\right] \bigg\}\,,
\eea
where $r \equiv \beta_{0}\,\alpha_S$ has the role of regularization parameter of the Landau singularities. The functions $\widetilde{f}_n(\lambda)$ ($n=1,2$)
are analytic for any $0\leq\lambda< +\infty$ and satisfy the following conditions:
\beq
\widetilde{f}_n(\lambda)|_{r=0} \, = \, {f}_n(\lambda)\,,\qquad 
\widetilde{f}_n(\lambda=0) \, = \, f_n(\lambda=0)=0\,.
\eeq
This ``tilde'' regularization can be extended
to higher orders.
However, let us observe that the function 
$\widetilde{f}_1(\lambda)$ can be directly obtained from the function ${f}_1(\lambda)$ by means of the following
substitution:
\bea
\label{subreg}
&&\!\!\!\!\!\!\!\!\!\!\!\!\!\!\!(1-k\lambda)\ln(1-k\lambda)
\, \mapsto \,
\frac{1}{2} \Bigg\{(1-k\lambda)
\ln\left[(1-k\lambda )^2+r^2\right]-\ln(1+r^2)\\ \nn
\!\!\!\!\!\!\!\!\!\!\!\!&&\!\!\!\!\!\!\!\!\!\!\!\!-\frac1r\left[
\left( (1-k\lambda)^2-r^2 \right) \arctan\left(\frac{1-k\lambda}{r}\right)-(1-r^2) \arctan\left(\frac{1}{r}\right) + \pi\lambda^2 \right]\Bigg\}\,,\,\,\, k=1,2\,.
\eea
Actually, one can use the above substitution 
to obtain an approximation 
of the higher-order resummation functions
$\widetilde{f}_n(\lambda)$, $n \ge 3$\,%
\footnote{At NLL, the substitution in Eq.\,(\ref{subreg}) produces a
different function with respect to the 
$\widetilde{f}_2(\lambda)$ of Eq.\,(\ref{f2new}). 
However, we have checked that, even though
the functional forms of the two functions are different,
the numerical differences between them
are negligible.}.
Therefore the corresponding form factor, being free from Landau singularities, can be integrated either 
exactly (in numerical way), or evaluated analytically by means of the saddle-point method.
That allows us a cross-check for the inversion of the resummed form factor, 
as well of course as an estimate of the accuracy of the saddle-point method, even in the small limit $y \to 0^+$.


\section{Numerical results for the thrust resummed form factor}
\label{sec_pheno}

We now apply the results obtained in the previous sections to the case of the form factor for the thrust distribution in electron-positron annihilation.

The thrust $T$, a classic, infrared-safe shape variable, is defined as\cite{Farhi:1977sg}:
\begin{equation}
\label{thrust}
T 
\equiv 1-\tau = {\rm max}_{\bf n} \frac{\sum_i|{\bf p}_i \cdot {\bf n}|}{\sum_i |{\bf p}_i|}\,,
\end{equation}
where the sum is over all final-state particles $i$ with spatial momentum ${\bf p}_i$. The maximum is taken with respect to the direction of the unit three-vector ${\bf n}$.

\begin{figure}[ht]
\begin{center}
\includegraphics[width=0.7\textwidth]{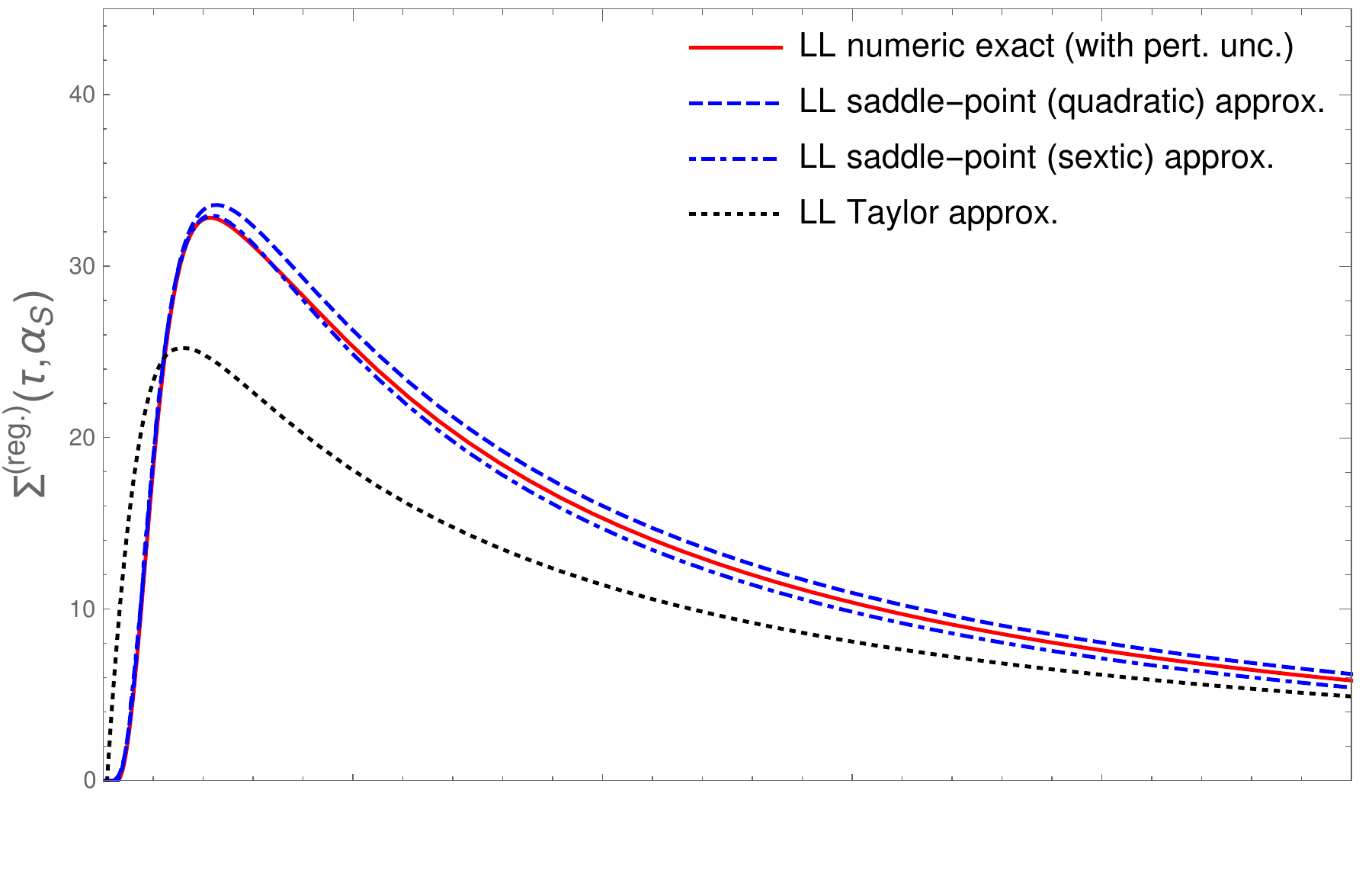}\vspace*{-.8cm}\\
\includegraphics[width=0.705\textwidth]{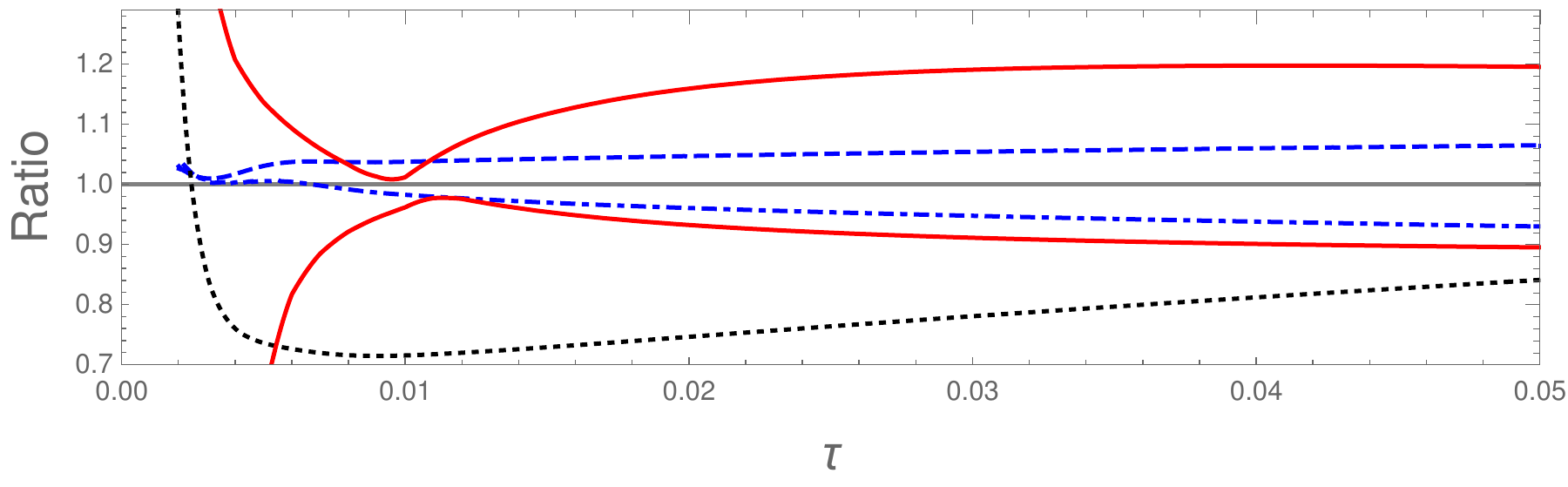}
\end{center}
\vspace*{-.5cm}\caption{
\label{figLL}
      {\em
Resummed form factor at leading logarithmic accuracy obtained through:
exact numerical inversion (red solid line); 
analytic saddle-point inversion (dashed blue line);  also including anharmonic sextic corrections (dot-dashed blue line) and
analytic Taylor inversion (black dotted line). 
The lower panel shows the ratios with respect to the exact numerical inversion and the band 
between the two red curves is the estimate
of the perturbative uncertainty coming from higher-order corrections.
}}
\end{figure}

The form factor for the thrust $\Sigma(\tau,\alpha_S)$ is strictly connected with
that of other shape variables such as the heavy-jet mass and more generally to the form factors
appearing in the context of threshold resummations. In particular the general structure of the form factor is universal, while at NLL and beyond
the explicit values of the numerical coefficients are observable dependent. The numerical results of this section are thus valid not only for
the specific case of the thrust but have a high degree of generality.

In Fig.\,(\ref{figLL}), we plot the resummed form factor at leading logarithmic accuracy for the thrust distribution at the center-of-mass energy $Q=m_Z=91.1876$\,GeV, obtained
by applying
the analytic procedure to the resummation function
${f}_1(\lambda) \to \widetilde{f}_1(\lambda)$. We show the exact numerical result (red solid line),
the saddle-point (quadratic or Gaussian) approximation (blue dashed line), the saddle-point approximation with the anharmonic corrections up to the sextic term (blue dot-dashed line) and the result obtained
with the standard analytic inversion based on a Taylor expansion (black dotted line).

\begin{figure}[t]
\begin{center}
\includegraphics[width=0.7\textwidth]{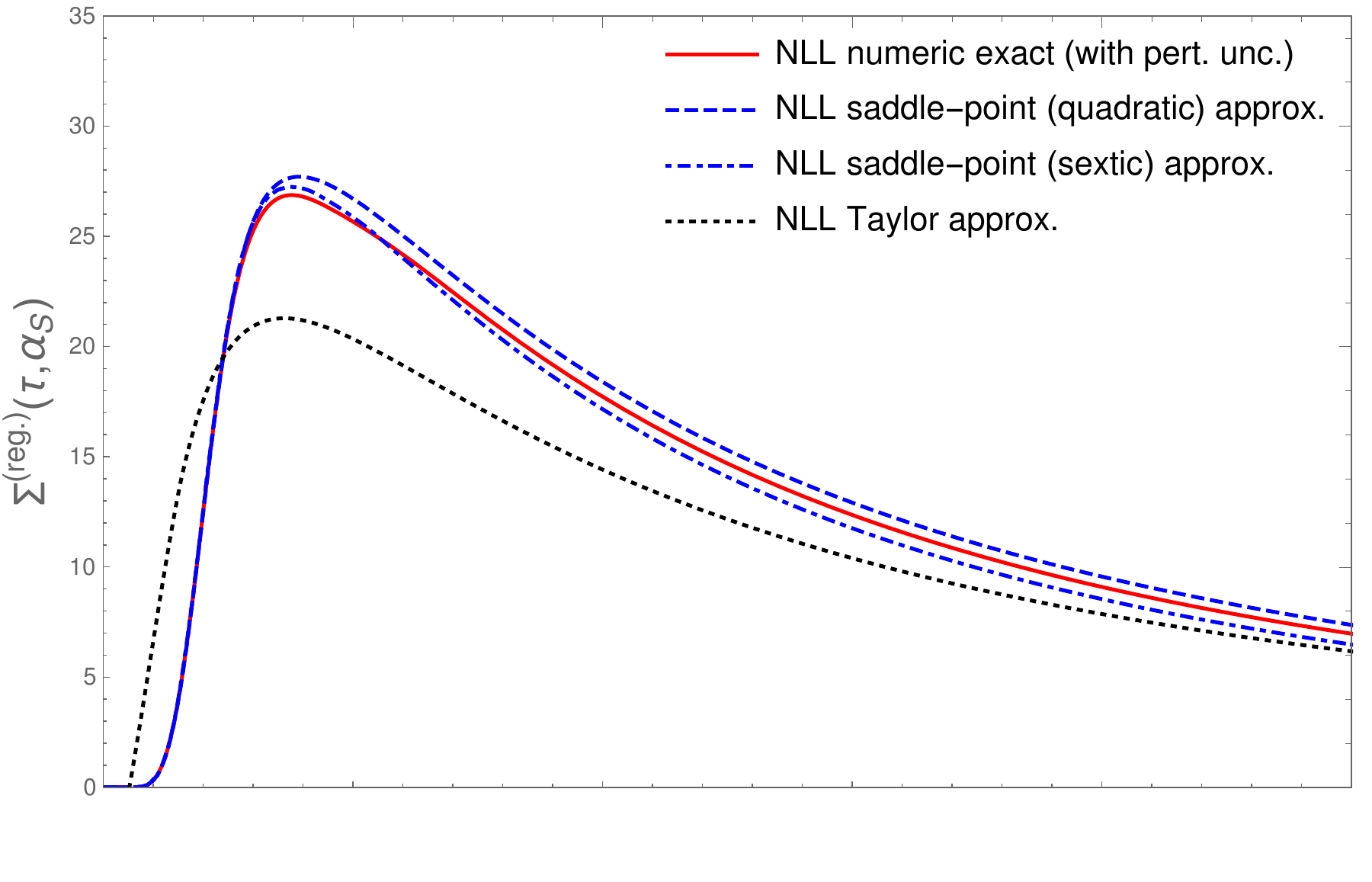}\vspace*{-.8cm}\\
\includegraphics[width=0.705\textwidth]{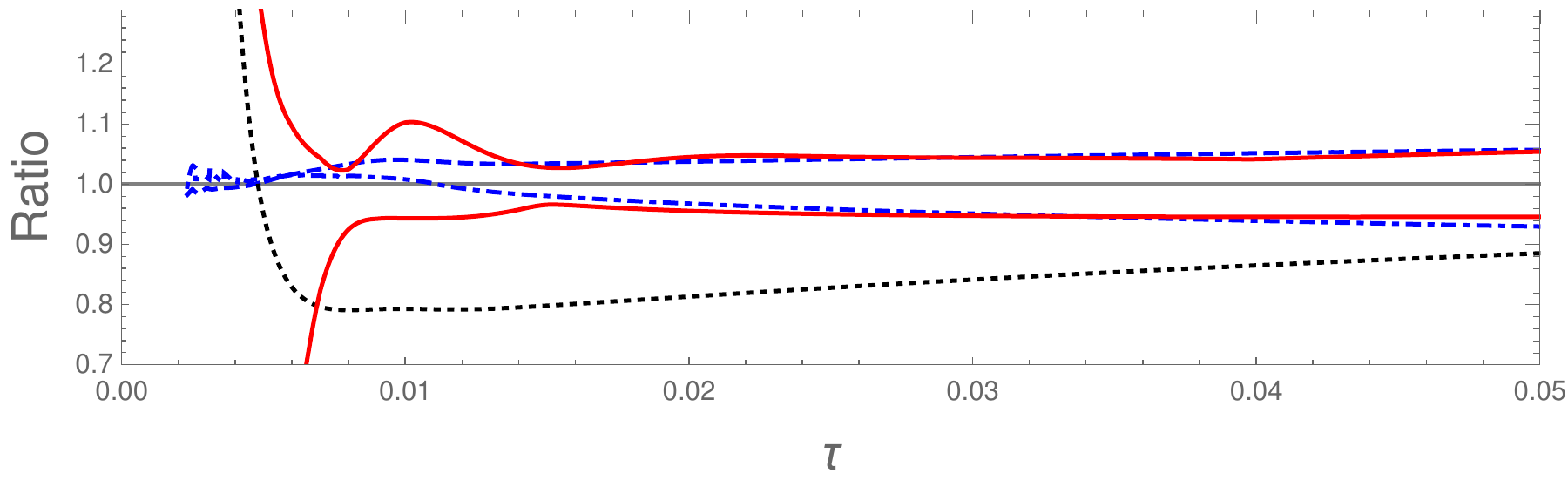}
\end{center}
\vspace*{-.5cm}
\caption{
\label{figNLL}
{\em Resummed form factor at next-to-leading logarithmic accuracy obtained through:
exact numerical inversion (red solid line); analytic saddle-point inversion (dashed blue line);  including also anharmonic sextic corrections (dot-dashed blue line) and
analytic Taylor inversion (black dotted line). The lower panel shows the ratios with respect to the exact numerical inversion and the band between the two red curves is the estimate
of the perturbative uncertainty coming from higher-order corrections.
}}
\end{figure}

The lower panel of Fig.\,(\ref{figLL}) shows the ratios
with respect to the exact numerical result together with the perturbative uncertainty of the exact result estimated both with the renormalization scale ($\mu_R$) variation by a factor of
two around the central value $\mu_R=Q$ (red solid lines) and through the impact of the next order corrections.
We observe that the
results obtained with the saddle-point method are in very good (percent level) agreement with the exact numerical result up to very low values of $\tau$, $\tau\sim 0.002$, where the cross section
is rapidly approaching zero. 
We also observe that the uncertainty of the saddle-point approximation is generally much smaller than the perturbative uncertainty.
On the contrary, the form factor obtained through the Taylor approximation is outside the perturbative uncertainty in a  wide range of $\tau\gtrsim 0.005$. This behavior is a consequence
of the fact that the resummation of the leading logarithms in momentum space {\itshape does not} correspond to the resummation accuracy of the leading logarithms in the conjugate
moment space $\ell\equiv \ln (1/\tau) \neq L\equiv \ln N$.

\begin{figure}[t]
\begin{center}
\includegraphics[width=0.7\textwidth]{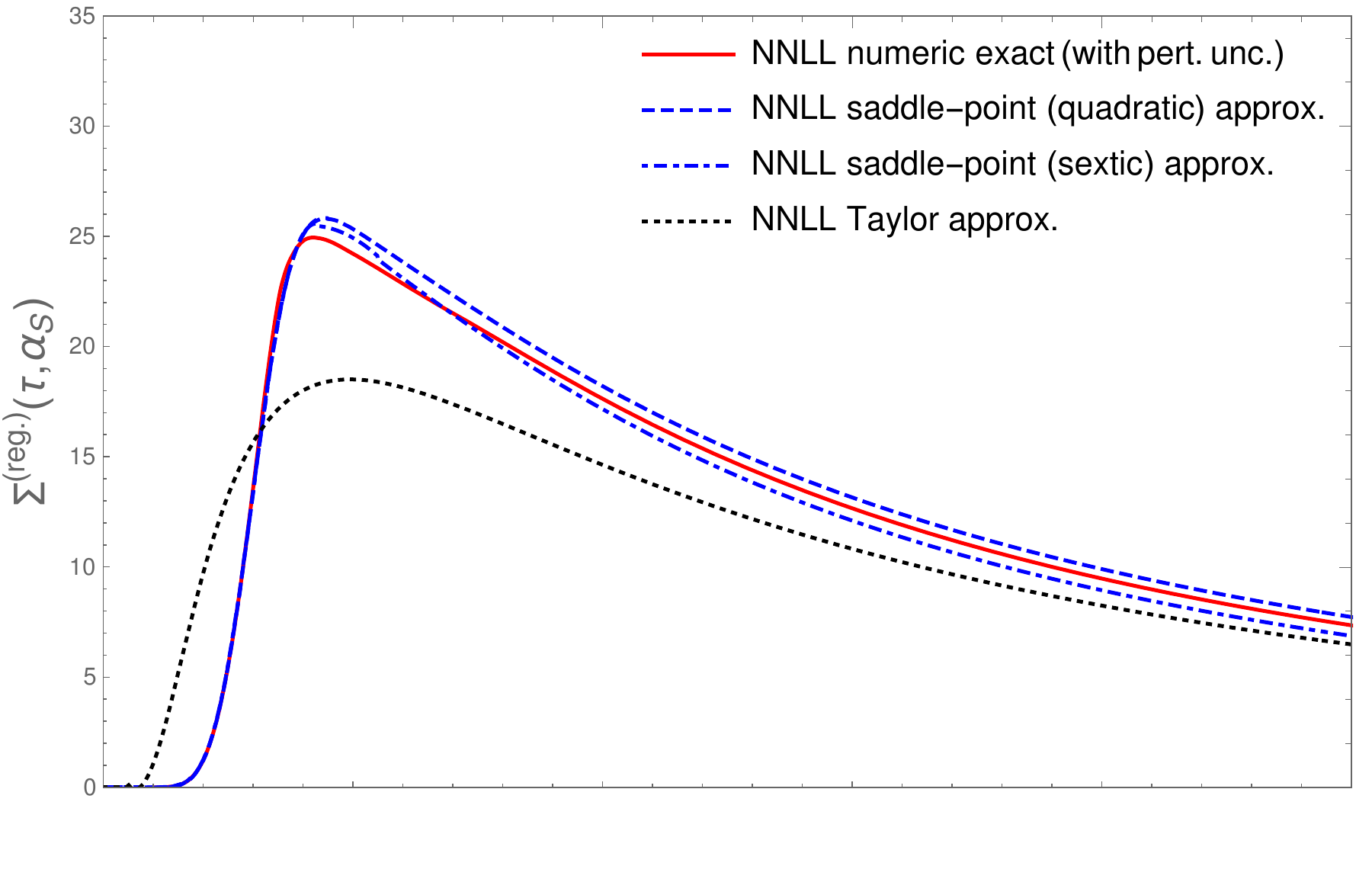}\vspace*{-.8cm}\\
\includegraphics[width=0.705\textwidth]{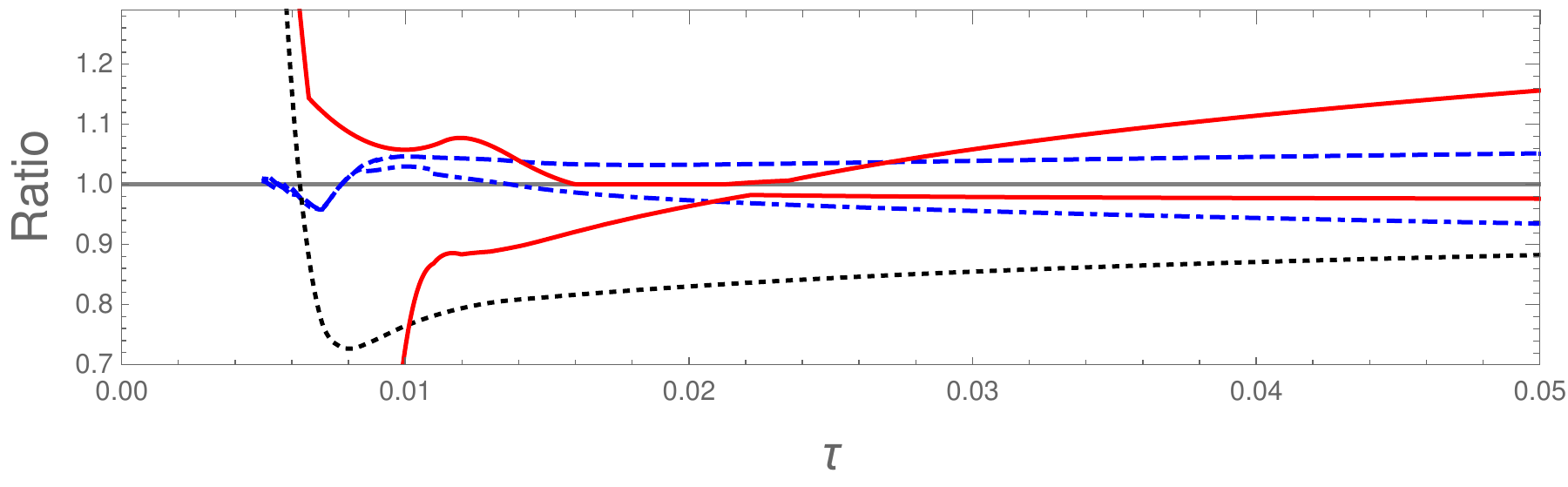}
\end{center}
\vspace*{-.5cm}
\caption{
\label{figNNLL}
      {\em
                Resummed form factor at next-to-next-to-leading logarithmic accuracy obtained through:
exact numerical inversion (red solid line), analytic saddle-point inversion (dashed blue line)  including anharmonic sextic corrections (dot-dashed blue line) and
analytic Taylor inversion (black dotted line). The lower panel shows the ratios with respect to the exact numerical inversion and the band 
between the two red curves is the estimate
of the perturbative uncertainty coming from higher-order corrections.
}}
\end{figure}

Fig.\,(\ref{figNLL}) and Fig.\,(\ref{figNNLL}) show the same results presented in Fig.\,(\ref{figLL})
at NLL and NNLL accuracy respectively. The results are both qualitatively and quantitatively very similar to the LL case.
This is not unexpected since the inversion methods (and their intrinsic accuracy) are independent from the logarithmic order of the perturbative resummed form factor.

We add some quantitative comments
on the sensitivity from the Landau singularity of the standard (unregularized) form factor in the case of the Taylor and the saddle-point approximated analytic inversion methods.
In the case of the Taylor expansion, we observe that there is no sensitivity from the Landau singularity.
This is because, as already discussed, the expansion point
$N=1/\tau$ is not the saddle point of the full theory but the one of the free theory which is insensitive to the Landau pole (which is a running coupling effect).
However this is also the reason for the lower accuracy of such inversion.
Conversely, in the case of the saddle-point inversion, 
instabilities arise at very small $\tau$. 
These instabilities are related to the singularities of the resummed functions $f_k(\lambda)$ in Eq.\,(\ref{f12}) starting at $N=N_L\sim  1000$ (which originate from the
Landau singularity of the standard QCD coupling).
At LL, we find that the saddle point
 $\xbar{N}$ rapidly increases at small $\tau$ reaching its maximum of $\xbar{N}\sim 800$ for $\tau \sim 0.004-0.006$.
 For lower values of $\tau$ no saddle-point solution exists and the saddle-point inversion
 method is not applicable without a regularization of such singularities. This is exactly the region where NP corrections are expected to become essential. Similarly no saddle-point
 solution exists at NLL and NNLL for $\tau\lesssim 0.006-0.01$. Therefore, thanks to its higher accuracy,  the saddle-point method shows a sensitivity to the singularities of the
 original $N$-space formula, naturally setting the minimum value of $\tau\sim \Lambda_{QCD}/Q$
 at which NP effects are not negligible anymore.

 Finally, we have checked that for the unregularized form factor the saddle-point method, when applicable (i.e. in the perturbative region), is fully consistent with the numerical inversion
 within the Minimal Prescription of Ref.\,\cite{Catani:1996yz}. We have thus been able to confirm, in a fully analytic framework, the numerical results obtained in Ref.\,\cite{Aglietti:2025jdj}.


\section{Conclusions}

In this work we have studied the form factor occurring in the resummation formalism for event-shape distributions in the two-jet (or Sudakov) region. We have presented a simple
and fully analytic formula for the inverse transform of the form factor, from the conjugate moment space (where phase-space factorization and thus exponentiation hold), to the physical
momentum space, where experimental data are collected.

Our formula is based on a local expansion around the saddle point of the interacting theory, while standard analytical inversions used in the literature are
based on a Taylor expansion around the free-theory saddle point.

The ``true'' saddle point is the solution of a complicated
transcendental functional equation.
We have determined it both numerically  by standard methods and analytically by means of a recursive method.
The analytical solution converges rapidly (within a few iterations) to the exact numerical result.

To avoid the Landau singularity of the QCD coupling in the infrared region,  we regularized the resummed form factor 
by replacing the standard (infrared divergent) QCD coupling with an analytic effective coupling free from the Landau singularity.

We have also considered the frozen coupling limit for the resummed form factor and we
identified the (large) expansion parameter formally controlling the convergence of the saddle-point expansion.

In the case of the thrust event-shape variable, we have shown that our analytic formula is in very good agreement with the exact (numerical) evaluation of the inverse
transform of the form factor, while it differs significantly from the
standard analytical inversion used in the literature. 
We have shown illustrative numerical results for the form factor in physical space up to next-to-next-to-leading logarithmic accuracy.

Thus, we have been able to confirm, within a full analytic approach, the results presented in Ref.\,\cite{Aglietti:2025jdj}.

\vspace{1cm}
\centerline{\bf Acknowledgments}
\vspace{.2cm}
We would like to thank Stefano Forte and Luca Guido Molinari for useful discussions.
~\\

\appendix
\section{Frozen coupling resummed form factor}
\label{sec_frozen_coupling}

In the frozen-coupling limit,
the 
the function $g_y(N)$ in Eq.\,(\ref{gy}) simplifies to 
\beq
g_y(N) = 
y \, N  -  \frac{A}{2} \, \ln^2(N)
 -  B \, \ln(N)\,,
\eeq
with $A=0.2$ 
which ensures the convergence
of the improper integral at infinity.
Through the change of variable
$N \mapsto M$ with:
\beq
\label{NM}
N  = \frac{A}{y}\ln\left(\frac{1}{y}\right) \, M= \frac{A \, \ell}{y} \, M\,,
\eeq
the frozen-coupling resummed form factor becomes
\bea
\label{sigmaFC}
\Sigma(y) &=&
\frac{A \, \ell}{y} \,
\exp\left\{ 
- \, \frac{A}{2} \, \ell^2
\left[ 1 \, + \, \frac{\ln(A \, \ell)}{\ell} \right]^2 
\, - \, B \, \ell
\left[ 1 \, + \, \frac{\ln(A \, \ell)}{\ell} \right]
\right\} \times
\nonumber\\
&\times&
\int_{c-i\infty}^{c+i\infty} 
\frac{dM}{2\pi i} 
\, \varphi(M)
\, \exp\Bigg(
A \, \ell 
\left\{ 
M \, - \left[ 
1 \, + \, \frac{\ln(A \, \ell)}{\ell} \right]\ln\left(M\right)
\right\}
\Bigg)\,,
\eea
where we have defined the 
function
\beq
\varphi(M) \, \equiv \, 
\exp\left[ - \, \frac{A}{2} \, \ln^2(M)
\, - \, B \, \ln(M)
\right]\,.
\eeq
We have split the terms at the exponent
into three different classes:
\begin{enumerate}
\item
Terms independent on the integration variable $M$, to be taken
outside the integral;
\item
Terms dependent on $M$, but
independent on $\ell \gg 1$, which are relegated
into the function $\varphi(M)$;
\item
Terms dependent on both $\ell$ and $M$,
which are written by factoring out the
term $A\ell$. 
\end{enumerate}

In order to justify the convergence of the
saddle-point expansion, we consider
the limit
\beq
\ell \, \mapsto \, + \, \infty,
\eeq
which is implied by the two-jet kinematic limit
\beq
y \, \mapsto \, 0^+.
\eeq
By means of the change of variable 
$N \mapsto M$ in Eq.\,(\ref{NM}),
we have been able to relegate most
of the dependence on the large parameter
$\ell$ in the coefficient
of the curly bracket of Eq.\,(\ref{sigmaFC}).
Indeed, the residual dependence 
of the exponent of the integrand on $\ell$
involves the power correction
\beq
\frac{\ln(A \, \ell)}{\ell},
\eeq
which vanishes for $\ell \to \infty$.
%
Therefore, 
we can basically identify the parameter $\ell$
with the auxiliary
parameter $s \gg 1$ controlling the saddle-point approximation in Eq.\,(\ref{eq_saddle_point}).

\noindent
In the saddle-point (Gaussian) approximation, we finally obtain:
\beq
\label{sigmaFCsp}
\Sigma(y)  \simeq  \frac{A \ell}{y}
\exp\left( - \, \frac{A}{2} \ell^2 K^2 \, - \, B \ell K \right)
\, \varphi(K) \,
\sqrt{ \frac{K}{2 \pi A \, \ell} } \,
\left( \frac{e}{K} \right)^{A \, \ell \, K}\,,
\eeq
with
\beq
K 
\equiv \, 1 \, + \, \frac{\ln(A\ell)}{\ell}\,.
\eeq

\begin{figure}[ht]
\begin{center}
\includegraphics[width=0.7\textwidth]{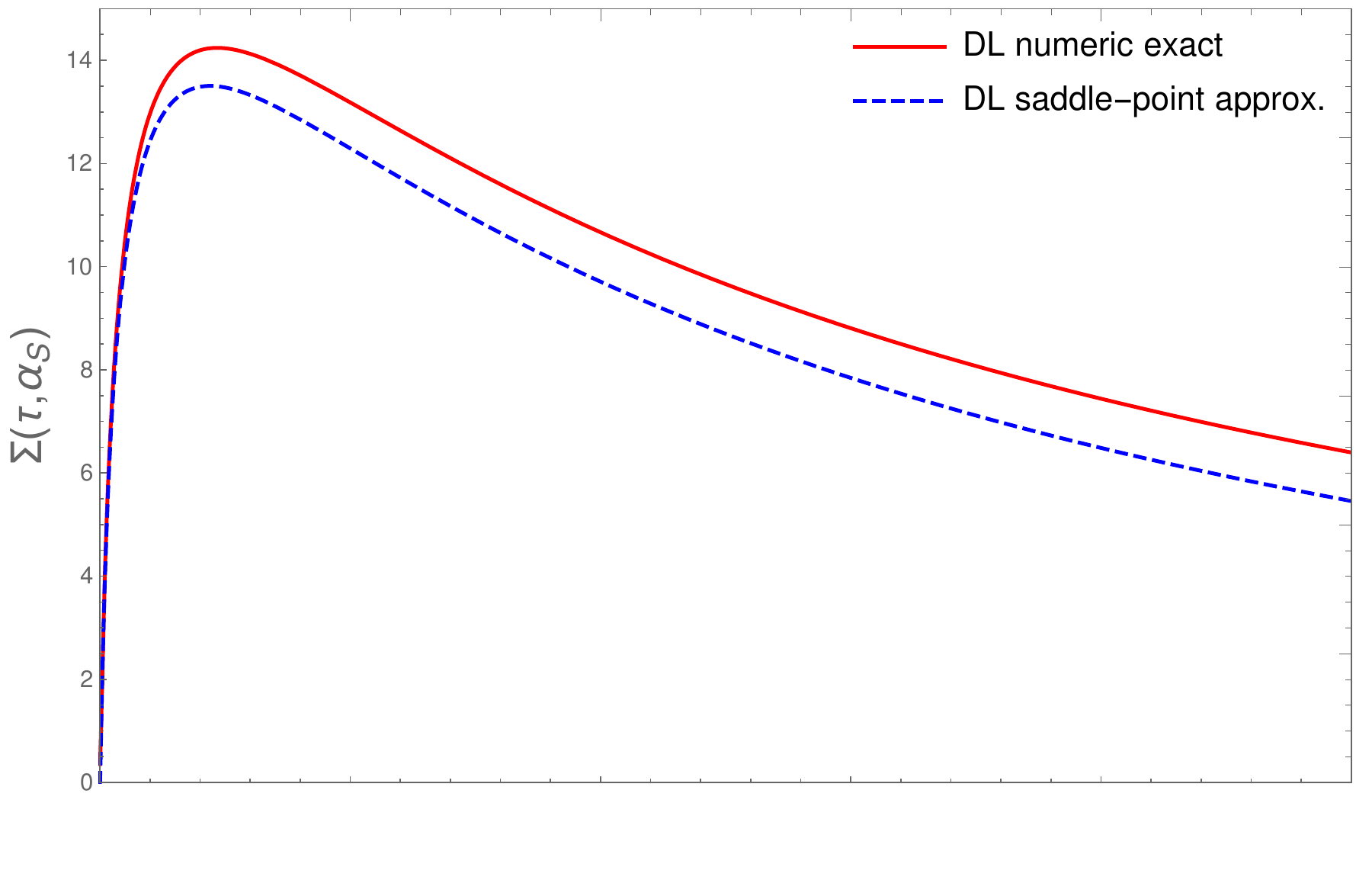}\vspace*{-.8cm}\\
\hspace*{-.2cm}\includegraphics[width=0.715\textwidth]{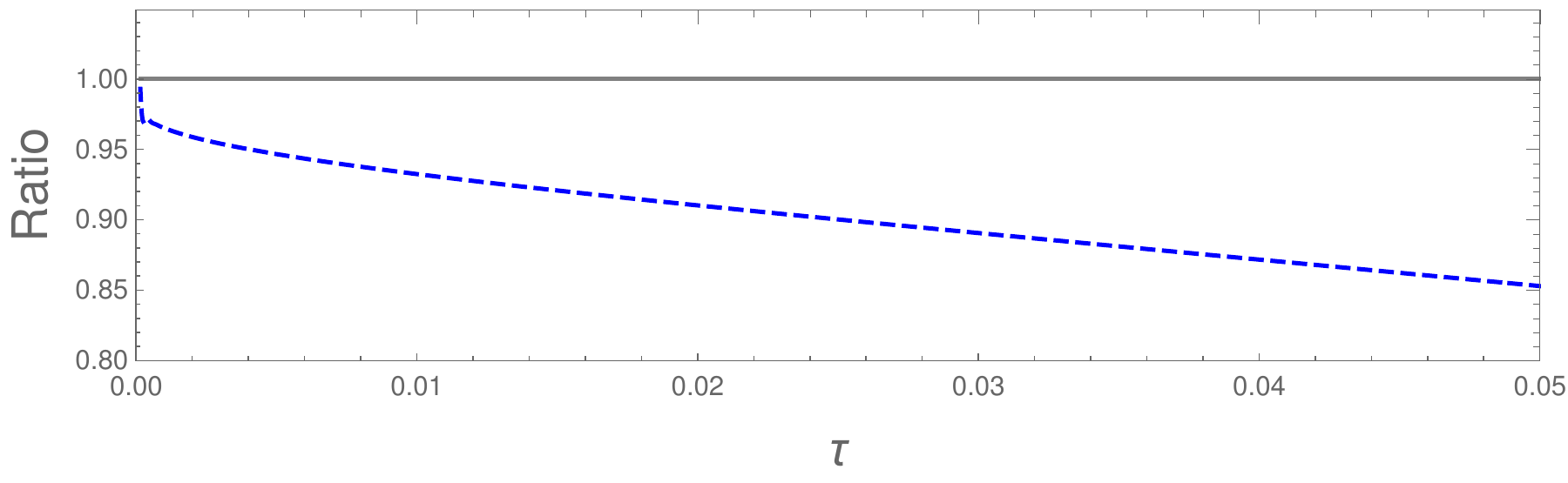}
\end{center}
\vspace*{-.5cm}
\caption{\em 
  Exact (numeric) frozen-coupling resummed form factor, Eq.\,(\ref{sigmaFC}), (red solid line) compared with the saddle-point result, Eq.\,(\ref{sigmaFCsp}), (blue dashed line).  
\label{figDL}
{\em
}}
\vspace*{.5cm}
\end{figure}

In Fig.\,(\ref{figDL}), we show the thrust resummed form factor in the frozen coupling limit at double logarithmic accuracy (
$A= 0.2$ and $B=0$).
We show the exact numerical result (red solid line) compared with
the saddle-point approximation
(blue dashed line)\,.
As expected the accuracy of the saddle-point approximation increases in the back-to-back limit $\tau\to 0$ where $\ell\to\infty$.

Finally, since the time-like coupling $\widetilde{\alpha}_S(q^2)$ introduced in Sec.\,\ref{subsec_reg}
saturates
at very small values of the squared momentum
transfer $q^2$,
\beq
\lim_{q^2\to 0} \widetilde{\alpha}_S\left(q^2\right)
\, = \, \frac{\pi}{\beta_0} \, \simeq \, 1.6,
\eeq
we expect the frozen-coupling limit
to actually provide a good approximation
to the full (regularized) form
factor 
for very small $y$.
%
We can then claim
that also in the running coupling case 
the saddle-point expansion is justified
for $y \ll 1$, as indeed we have numerically shown in Sec.\,\ref{sec_pheno}.

\end{document}